\documentclass{aa501}
\usepackage{graphicx}
\begin{document}
   \title{New infrared star clusters and candidates in the Galaxy detected with 2MASS}

   \author{C.M. Dutra\inst{1,2} \and E. Bica
          \inst{1}}
 
   \offprints{C.M. Dutra -- dutra@iagusp.usp.br}
 
  \institute{Universidade Federal do Rio Grande do Sul, IF, 
CP\,15051, Porto Alegre 91501--970, RS, Brazil\\
 \and
 Instituto Astronomico e Geofisico da USP, CP\, 3386, S\~ao Paulo 01060-970, SP, Brazil\\
}

   \date{Received ; accepted }
 
\abstract{
A sample of 42 new infrared star clusters, stellar groups  
and candidates was found throughout the Galaxy in the 
2MASS J, H and especially K$_S$ Atlases. In the  
Cygnus X region 19 new clusters, stellar groups and candidates were 
found as compared to 6 such objects in the previous  literature. Colour-Magnitude Diagrams using the 2MASS 
Point Source Catalogue provided preliminary distance estimates in the range  
1.0 $<$ {\it d$_{\odot}$} $<$ 1.8 kpc for 7 Cygnus X clusters. Towards the central parts of the Galaxy 7 new IR clusters 
and candidates were found as compared to 61 previous objects. A search
for prominent dark nebulae in K$_S$ was also carried out in the central bulge area. We report  5 dark nebulae, 2 of them are candidates for molecular clouds able to generate  massive star clusters near the Nucleus, such as the Arches and Quintuplet clusters.
}

\maketitle

%
 
\section{Introduction}
 
The digital Two Micron All Sky Survey (2MASS)  infrared atlas  (Skrutskie et al. \cite{skru} -- web interface {\it http://www.ipac.caltech.edu/2mass/}) 
can provide  a wealth of new objects for  future studies with large telescopes, 
a role similar to that played in the optical by the Palomar and ESO/SERC Schmidt 
plate Atlases  in past decades. Regarding Galactic IR star clusters two basic 
classes can be identified in the 2MASS survey: (i) resolved embedded clusters in 
nearby complexes such as those in Perseus, Orion and MonR2 which were studied by means of star counts (Carpenter \cite{carp1}), and
(ii) partly resolved clusters or candidates located in more distant parts of the Galaxy, such
as towards the Centre (Dutra \& Bica \cite{db20}, hereafter Paper I).

As the IR  search for new objects  progresses, it is worth recalling
how the development of overall optical catalogues of similar objects
occurred. For comparison purposes,  no survey or search of optical star clusters and  nebulae has ever been complete.  The samples currently available in overall optical catalogues  arise from many contributions, and in the following 
we mention some examples. The open cluster catalogues as compiled by Alter et 
al. (\cite{alt}) and Lyng\aa~(\cite{lyng}) had major contributions from short communications including small lists of newly found objects. Despite a systematic search for faint clusters on the Palomar plates like that generating the Berkeley clusters (Setteducati \& Weaver \cite{set}), or  the individual lists which generated the ESO catalogue (Lauberts \cite{lau} and references therein), until quite recently new star clusters and candidates both on the Palomar and ESO/SERC Schmidt plates could still be  found (e.g. the 6 new objects in Saurer et al. \cite{sau}). All catalogues of open stellar clusters are biased in one way or the other, and their statistical
value is uncertain. Concerning optical nebulae, the northern/equatorial Sh2 catalogue (Sharpless \cite{sha}) had  important complementary contributions,
e.g. the BFS objects (Blitz et al. \cite{bfs}). In the Magellanic Clouds the same process has taken  place over decades (e.g. Bica \& Dutra \cite{bic3}, Bica et al. \cite{bic2} and references therein).

In the present study we make use of the 2MASS survey in the J (1.25$\mu$m), H (1.65$\mu$m) and K$_s$ (2.17$\mu$m) bands to search for new  IR star clusters in the Galaxy, in particular towards the  Cygnus X region. In  Sect. 2 we discuss the search criteria and present the results. In Sect. 3 we discuss the angular distribution of the sample
and  preliminary implications  of the cluster detections, as well as of some newly found dark and bright nebulae. Finally, the concluding remarks are given in Sect. 4.

\section{Search for IR star clusters and candidates in the 2MASS Atlas}
 
The search was carried out in selected 2MASS areas with
evidence of star formation, while that of Paper I was systematic  in a specific
area. The selection follows procedures for cross-identifications and new 
detections
of extended objects as outlined in Bica \& Schmidt (\cite{bic}) and Bica
et al. (\cite{bic2}).

 The  JHK$_s$ images  were obtained by means of the 2MASS Survey Visualization \& Image Server facility in the web interface {\it http://irsa.ipac.caltech.edu/}. The K$_s$ band allows one to probe  deeper in more absorbed regions, and the J and H bands were used mostly as control images for the presence of bright stars, and as comparisons to estimate  how reddened the objects could be, and what to expect in optical bands.  We also checked for possible optical counterparts using 1st (DSS) and 2nd (XDSS) generation digitized sky survey extractions. 
In the generated sky charts we were also guided by the presence of IRAS sources in general (Beichman et al. \cite{bei}) and/or ultracompact H\,II regions (Kurtz et al. \cite{kur}). We also checked the objects for possible relationship to dark nebulae (Lynds \cite{lyn}). 
For some objects the absorption is patchy even in the $K_s$ band, so that one cannot rule out the possibility of an enhancement of field stars -- a small dust window.
 
The present work was not exhaustive and it is not intended to be complete in 
any sense in the currently available 2MASS second incremental release Atlas areas. Nevertheless, some preliminary detection rates can be inferred  in view of future searches. {We focused our attention mainly on the Cygnus X area, but we also continued the exploration of central directions in the Galaxy, as well as some other samplings along the disk.}

\subsection{Targets in Cygnus X}

In the Cygnus X area we concentrated efforts on the large angular size 
continuum sources (Downes \& Rinehart \cite{dori}, Wendker \cite{wen}), thus avoiding many small discrete sources  related to the thermal emission fine structure (Wendker et al. \cite{wen3}). Indeed, 
optical H\,II regions often show embedded clusters which are small with respect to the overall
emission extent and detailed structure of the complex. Out of the 21 DR sources 11 are  now known to harbour a star cluster or stellar aggregate (Sect. 3.1). 
Wendker's (\cite{wen}) continuum sources basically include those in  Downes \& Rinehart (\cite{dori}), and several of the additional clusters and aggregates  in the area (Table 1, Sect. 3.1) are related to them. Excluding those related to DR sources, the cluster and aggregate detection rate is $\approx$1:4 for the remaining Wendker (\cite{wen}) sources, which in turn suggests that most of the remaining radio continuum sources are related to the details of the continuum 
distribution, like thermal shells from individual massive stars, ridges and some non-thermal sources. Alternatively, some non-detections might be due to high 
K$_s$ absorptions for deeply embedded clusters in dense molecular cores as seen
in isolated relatively high Galactic latitute complexes  (Dutra \& Bica \cite{db20b}
and references therein).

\subsection{Targets in central parts of the Galaxy}

Towards  the central parts of the Galaxy we based our search mostly on the 
radio continuum catalogue by Kuchar \& Clark (\cite{kuc}), and the radio recombination line catalogues   by  Caswell \& Haynes (\cite{cas}) and Lockman (\cite{loc}). For such directions a preliminary estimate suggests that stellar enhancements in the K$_s$ band are detectable for 1:5 radio HII\,regions. This might be 
explained by important  intervening dust contributions for such directions (Dutra \& Bica \cite{db20b}), combined to the internal absorption arising from the cluster dust complex itself. Star crowding is an additional difficulty as compared to the Cygnus X region.

\subsection{Targets along the Disk}

Towards the anticentre and southern Milky Way the search was mostly aimed at a 
few specific entries from optical nebula catalogues (Sharpless et al. \cite{sha}, Rodgers et al. \cite{rcw}, Blitz et al. \cite{bfs}). Some of these H\,II regions or complexes have known optical open clusters, e.g. Tr\,14, Tr\,16 and their neighbours in the $\eta$ Carinae Complex (Feinstein \cite{fei}, V\'azquez et al. \cite{vaz} and references therein), or the IR  massive  central  star cluster in the RCW\,38 Complex  as recently shown in detail by means of  VLT/ISAAC imaging (Moorwood et al. \cite{moor}). In the present study  one additional cluster was detected in each of these optical complexes (Sect. 3.3). Perhaps the best example of detection possibilities in optical H\,II regions is the Sh2-254/258 Complex. Five prominent discrete H\,II regions led to 4 IR clusters, only one was previously reported (Sect. 3.3).

\subsection{Results}
 
  A total of 42 newly detected IR star clusters, stellar groups and candidates are given in Table 1. By columns: (1) a running number identification, (2) and (3) Galactic coordinates, (4) and (5) J2000.0 equatorial positions, (6) and (7) major
and minor angular dimensions, and (8) remarks such as degree of resolution, classification, duplicity, and related nebulae and other sources. We distinguish star clusters from less populous and/or less dense possible physical systems, which are referred to as stellar groups. Angular diameters of the objects in Table 1 are typically 1$^{\prime}$--2$^{\prime}$, thus suitable for deep images in large telescopes.
 
We illustrate in Fig. 1 the object 13 from Table 1, which is the IR resolved star cluster in the HII
region DR\,22 in Cygnus X. This cluster is surrounded by a related sparse group of stars (object 14).
Recently an IR stellar group or aggregate was discussed in the HII/photodissociation region DR\,18 in Cygnus X (Comer\'on \& Torra \cite{com}). A newly found IR nebula ($\alpha$(2000) =  20$^h$ 39$^m$ 25.9$^s$,$\delta$(2000) = 41$^{\circ}$ 20$^{\prime}$ 02$^{\prime\prime}$) is also present  in Fig. 1. In this nebula no candidate cluster was detected, since in the 2MASS J image a single bright star is centred in the nebula. A similar IR nebula was studied by Comer\'on \& Torra (\cite{com}) in DR\,18. The candidate clusters in Table 1 are predominantly unresolved objects and/or much embedded in nebulae in the IR bands. Fig. 2 illustrates the case of Object 40, which is related to the radio H\,II region G353.4-0.4.

 \begin{figure}
   \centering
 \resizebox{\hsize}{!}
 {\includegraphics{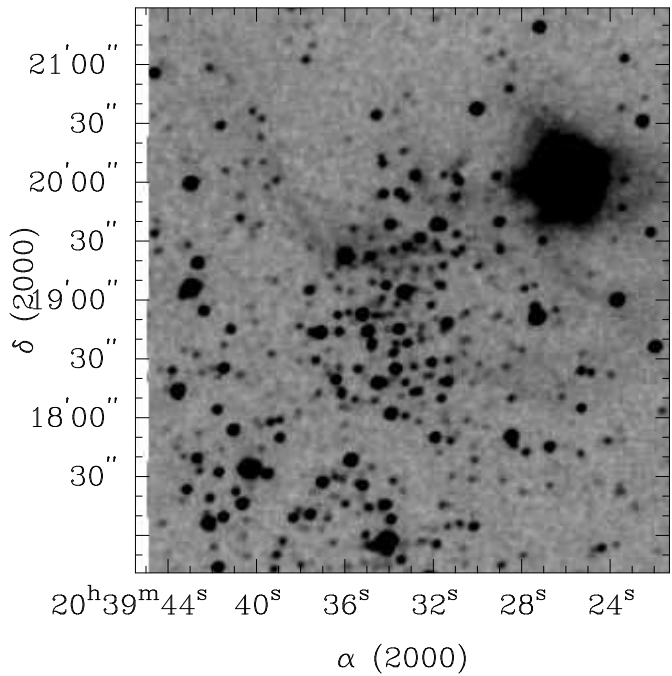}}
\caption{5$^{\prime}$ $\times$ 5$^{\prime}$ 2MASS K$_s$ image of the IR cluster ($\alpha$ =  20$^h$39$^m$34$^s$ and $\delta$ = 41$^{\circ}$19$^{\prime}$05$^{\prime\prime}$ J2000) in the radio HII region DR\,22. A newly identified IR compact nebula is located at $\approx 2^{\prime}$ to the northwest}
\label{FigGam}
\end{figure}

\begin{figure}
\centering
 \resizebox{\hsize}{!}
 {\includegraphics{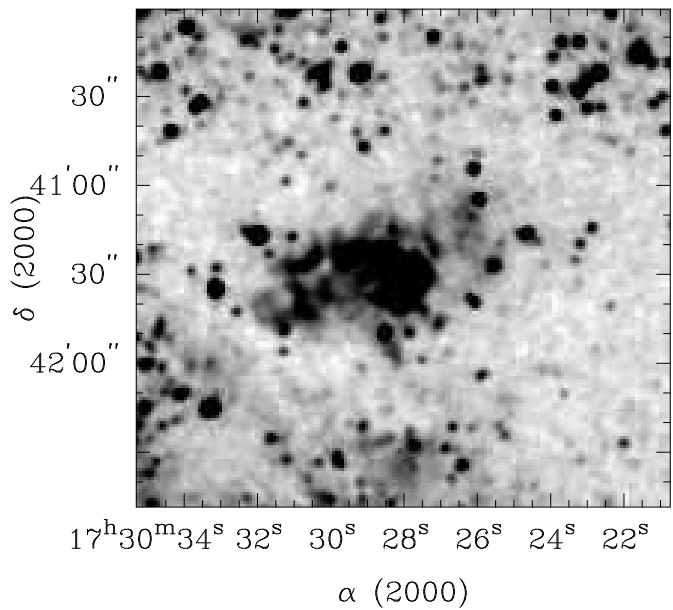}}
\caption{$3^{\prime}\times 3^{\prime}$ 2MASS K$_s$ image of the predominantly unresolved cluster candidate ($\alpha$ =  17$^h$30$^m$28$^s$ and $\delta$ = -34$^{\circ}$41$^{\prime}$30$^{\prime\prime}$ J2000) in the radio HII region G353.4-0.4.} 
\label{FigGame}
\end{figure}  

\begin{table*}
\caption[]{New infrared star clusters, stellar groups or candidates}
\begin{scriptsize}
\label{tab1}
\renewcommand{\tabcolsep}{0.9mm}
\begin{tabular}{lrrrrrrl}
\hline\hline
Object$^{**}$&$\ell$&{\it b}& $\alpha$(2000) & $\delta(2000)$&D & d& Remarks$^*$\\
&($^{\circ}$)&($^{\circ}$)&hh:mm:ss.s &$^{\circ}$:$^{\prime}$:$^{\prime\prime}~$&$^{\prime}$&$^{\prime}$~~&\\
\hline
 
1& 2.20&   3.45& 17:37:40.7&-25:15:03&   1.9 &  1.7 &resolved cluster or dust window, related to LDN74? partly optical\\
2& 3.73&   3.96& 17:39:20.1 &-23:41:00&    1 &  0.6 &resolved stellar group, related to LDN114? partly optical\\
3&66.96&  -1.28& 20:04:52.9 & 29:11:45&    2 &   2 &mostly a resolved cluster in nebula IC4955=BFS2c, in complex  IC4954\\
4& 67.00&  -1.22 &20:04:45.4&  29:15:05&  0.5&  0.4& partly resolved stellar group in nebula BFS2a, in complex IC4954 \\
5& 75.77&   0.34& 20:21:41.8&  37:25:50&  1.7&  1.7& cluster and/or IR nebula in radio HII region G75.8+0.4,rel. to OH maser ON2,backgr. of cluster Be87 \\
6& 77.96&  -0.01& 20:29:36.9&  39:01:15&    3&    3& partly resolved cluster in radio HII region DR9 \\
7& 78.04&   0.62& 20:27:13.0&  39:26:56&  3.5&    3& resolved cluster in radio HII region DR6 \\
8& 78.16&  -0.37& 20:31:45.4&  38:58:00&    2 &   2& partly resolved cluster  in Radio HII region DR13 \\
9& 78.16&  -0.55& 20:32:27.8&  38:51:26&  1.4 & 1.2& partly resolved cluster related to IRAS20306+3841, in  radio complex W69 \\  
10& 79.30&   0.29& 20:32:29.0&  40:16:30&  2.5 & 2.0& mostly a resolved cluster in radio HII region G79.306+0.282 in DR15, related to IRAS20306+4005\\
11& 79.31&   1.30& 20:28:09.7&  40:52:42&  3.5&  3.0& mostly a resolved cluster in Radio HII region DR7, related to  IRAS20264+4042\\
12& 79.87&   1.18& 20:30:28.0&  41:15:48&  1.8&  1.8& resolved stellar group in radio HII region DR11, related to IRAS20286+4105\\
13& 80.94&  -0.16& 20:39:34.0&  41:19:05&  2.3&  1.7& resolved cluster in radio HII region DR22, in stellar group 14 \\
14& 80.94&  -0.16& 20:39:33.0&  41:18:46&    4&    3& resolved stellar group in radio HII region DR22, includes cluster 13 and a compact nebula\\
15& 81.31&   1.10& 20:35:23.0&  42:22:05&  2.9&  2.2& resolved cluster in radio HII region DR17 \\
16& 81.44&   0.48& 20:38:29.0&  42:06:25&  2.3&  2.0& mostly a resolved cluster, in radio complex W75, related to radio HII region G81.5+0.6? \\
17& 81.57&  -0.07& 20:41:15.9&  41:51:51&  1.7&  1.7& resolved cluster in radio HII region DR23 \\
18& 81.66&  -0.02& 20:41:20.1&  41:58:22&  1.8&  1.3& resolved stellar group at the edge of radio HII region  DR23, related to DR23? \\
19& 81.71&   0.58& 20:38:56.8&  42:22:45&    2&    2& partly resolved stellar group in radio HII region  W75, pair with W75S IR cluster\\
20& 82.00&   0.80& 20:38:37.0&  42:39:24&   1.5&  1.5& mostly a resolved cluster in radio HII region W75N, pair with W75N IR cluster \\
21& 82.04&   2.33& 20:32:21.8&  43:41:05&  2.5&  2.0& resolved cluster in radio HII region G82.0+2.3, related to radio HII region DR16? \\
22& 82.57&   0.40& 20:42:33.5&  42:56:50&  1.7 & 1.5& partly resolved cluster in radio HII region G82.6+0.4 \\
23& 83.94&   0.78& 20:45:37.6&  44:15:21&  1.8&  1.8& partly resolved stellar group  in radio HII region G83.941+0.781 \\
24&173.48&   2.45&  5:39:12.5&  35:45:53&  1.2&  1.2& partly resolved  cluster in HII region  Sh2-233SE, pair with Sh2-233SE IR Cluster\\
25&182.05&   0.42&  5:52:04.9&  27:23:32&   3.5&   3.5& resolved cluster or dust window, related to IRAS05489+2723?, partly optical \\
26&186.13&   2.59&  6:09:28.2&  24:55:40&    4&    3& resolved stellar group related to IRAS06063+2456, partly optical\\
27&189.23&   0.90&  6:09:31.0&  21:23:57&    2&    2& resolved stellar group related to IRAS06065+2124 \\
28&189.69 &  0.72&  6:09:46.9&  20:54:44&  2.5&  2.5& resolved stellar group related to IRAS06067+2055 \\
29&191.92&   0.82&  6:14:45.1&  19:00:28&  2.5&    2& resolved cluster in nebula BFS52, related to IRAS06117+1901 \\
30&192.54&  -0.15&  6:12:24.9&  17:59:28&  0.9&  0.9& partly resolved cluster in HII region Sh2-254 \\
31&192.60&  -0.12&  6:12:36.6&  17:56:55&  1.7&  1.7& mostly a resolved cluster in HII region Sh2-256 \\
32&192.72&   0.03&  6:13:28.0&  17:55:27&    2&    2& partly resolved cluster in HII region  Sh2-258 \\
33&194.93 & -1.20&  6:13:21.2&  15:23:56&    1&    1& mostly an unresolved cluster related to IRAS06104+1524A, pair with stellar group 34 \\
34&194.94&  -1.22&  6:13:16.0&  15:22:30&  1.3&    1& mostly a resolved stellar group related to IRAS06103+1523, pair with cluster 33 \\
35&218.74&   1.85&  7:08:39.0&  -4:19:07&  0.6&  0.4& mostly unresolved cluster in HII region IC466=Sh2-288 \\
36&267.72&  -1.09&  8:58:05.0& -47:22:40&  1.2&  1.2& partly resolved cluster in reflection nebula vdBH-RN26, in RCW38? \\
37&287.81&  -0.82& 10:45:54.0& -59:56:58&  1.5&  1.5& partly resolved cluster in reflection nebula vdBH-RN43, in $\eta$ Carinae Complex? \\
38&351.61&   0.17& 17:23:23.4& -35:53:44&  0.4&  0.3& mostly unresolved cluster in radio HII region G351.6+0.2 \\
39&352.86&  -0.20& 17:28:19.1& -35:04:15&  0.3&  0.3& mostly unresolved cluster candidate in G352.866-0.199 \\
40&353.42&  -0.36& 17:30:28.2& -34:41:30&  1.7&  1.5& mostly unresolved cluster in radio HII region G353.4-0.4\\
41&354.67&   0.48& 17:30:24.0& -33:11:15&    1&  0.7& partly resolved cluster in radio HII region G354.664+0.470 \\
42&359.28&  -0.25& 17:44:52.7& -29:40:48&    2 & 1.7& resolved cluster in radio HII region G359.3-0.3, dust window? \\ 
\hline
\end{tabular}
\end{scriptsize}
\begin{list}{}
\item  Notes: $^*$ designations and related object types are - BFS: HII regions and/or reflection nebulae (Blitz et al. \cite{bfs}), Sh2: HII regions (Sharpless 1959), RCW: HII regions (Rodgers et al. \cite{rcw}), DR: radio HII regions in Cygnus X (Downes \& Rinehart \cite{dori}), G: radio HII regions (see text), vdBH-RN:  reflection nebulae (van den Bergh \& Herbst \cite{van}), LDN: dark nebulae (Lynds \cite{lyn}), ON: OH maser (Norris et al. \cite{nor}, and references therein), IRAS sources (Beichman et al. \cite{bei}).
\item  $^{**}$ after this manuscript had been completed we learned about a paper on Cygnus X 
(Comer\'on \& Torra \cite{com1}) that covers partially the same objects. 
\end{list}
\end{table*}
 
\section{Discussion}
 
\subsection{Cygnus X}

The Galactic longitude distribution in Table 1 indicates many new objects around l$\approx$ 80$^{\circ}$
which corresponds to the Local Arm plunge into Cygnus. We have
extensively surveyed this region with 2MASS which corresponds to the  HII region/molecular
complexes collectively known as Cygnus X (Downes \& Rinehart \cite{dori}). The direction of Cygnus X is very complex. By increasing the angular resolution  
the number of discrete radio continuum sources increased from about 20 in 
Downes \& Rinehart's study to nearly 800 in Wendker (\cite{wen2}). Recombination lines (e.g. Piepenbrink \& Wendker \cite{pie}) show that most sources are related to the Local Arm, but owing to the tangent orientation, kinematical distances basically cannot be inferred for {\it d$_{\odot}$} $<$ 3 kpc. The association Cyg\,OB2 is projected near the centre of Cygnus X and a fundamental question is whether a coherent local  physical phenomenon relates all these structures or whether they are simply line-of-sight projections  (Wendker et al. \cite{wen3}, Odenwald \& Schwartz \cite{ode}, and references therein). Cyg\,OB2 itself is an unusually  rich and compact association (Kn\"odlseder \cite{kno}).  The depth structure of Cygnus X is not yet known in detail, but available estimates as reviewed by Odenwald \& Schwartz (\cite{ode}) point to significant effects (0.7 $<$  {\it d$_{\odot}$} $<$ 2.5 kpc or more).
 
\begin{figure}
   \centering
 \resizebox{\hsize}{!}
 {\includegraphics{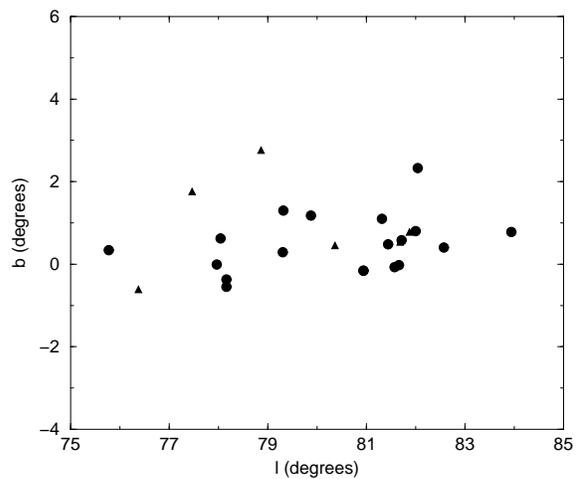}}
\caption{Cygnus X area: distribution of new IR clusters, stellar groups or candidates (circles) compared to those already known (triangles).}
\label{FigGa}
\end{figure}
 
 The angular distribution of  Table 1 objects in the Cygnus X area (Fig. 3) is more elongated in longitude than in latitude  (3:1). Assuming an average distance of 1.6 kpc, the objects are distributed along a region of $\approx$ 250 $\times$ 85 pc. This size is comparable to those of the largest OB associations, but the total number of 25 IR star clusters and stellar groups is remarkable. The new objects are 19, while previously known in the area were 6: (i) W\,75S (or DR\,21) IR cluster (Hodapp \cite{hod}), (ii) W\,75N IR cluster (Moore et al. \cite{moo}, Hodapp \cite{hod}); (iii) Sh2-106 IR cluster (Hodapp \& Rayner \cite{hod2}), (iv) IRAS\,20188+3928 IR Cluster (Hodapp \cite{hod}), (v) BD\,+40$^{\circ}$4124 stellar aggregate (Hillenbrand et al. \cite{hil}), (vi) DR\,18 stellar aggregate (Comer\'on \& Torra \cite{com}). 
 
  Embedded clusters can occur in pairs or triplets (e.g. NGC\,1333NE and NGC\,1333SW IR clusters -- Lada et al. \cite{lad}, and the subclusterings A, B and C of IRAS\,20050+2720 -- Chen et al. \cite{che}). Combining new and previous IR clusters and stellar groups in Cygnus X there occur 3 pairs: (i) IR cluster 17 and IR stellar group 18, (ii) W\,75S IR cluster and IR stellar group 19, and (iii) W\,75N IR cluster and IR cluster 20. Of 19 single and 3 double embedded IR clusters and stellar groups we derive a binary frequency of 14\% for Cygnus X.  This frequency probably reflects star formation and subsequent early dynamical evolution  from multiple molecular cores, in a potential dominated by dust and gas.

\begin{figure}
   \centering
 \resizebox{\hsize}{!}
 {\includegraphics{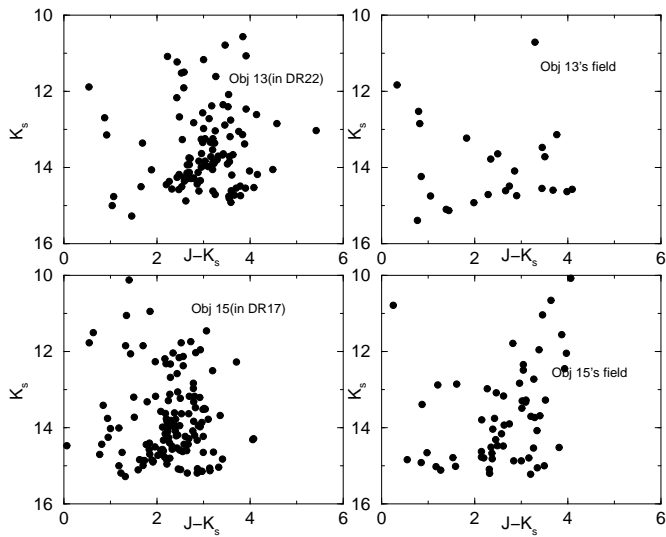}}
\caption{2MASS J $\times$ (J-K$_s$) colour-magnitude diagrams of the IR clusters in DR22 and DR17. Also shown are those for the equal area offset fields 3$^{\prime}$ to the South.}
\label{FigGa}
\end{figure}

Despite telescope aperture and seeing limitations with respect to the 
properties of Cygnus X clusters, the 2MASS Point Source Catalogue ({\it http://irsa.ipac.caltech.edu/applications/CatScan/}) can provide a preliminary diagnosis on depth effects.
We extracted all stars in equal areas for the cluster and offset field, 
and produced K$_s$ $\times$ (J-K$_s$) CMDs. According to the 2MASS  extractions the photometric errors are approximately $\epsilon$K$_s$ = 0.03 and $\epsilon$(J-K$_s$) = 0.05 at K$_s$ = 12.45  (typical 10th brightest star level). Close to the detection level at K$_s$ = 15.0 the errors increase to $\epsilon$K$_s$ = 0.18 and $\epsilon$(J-K$_s$) = 0.26. The apertures were circular with diameter equal to the major axis (Table 1). We illustrate in Fig. 4 the CMDs of the clusters in DR22 and DR17, together with their respective fields. A significant contrast occurs between the CMDs of cluster and field areas. The colour distribution in a cluster CMDs is wide (W(J-K$_s$) $\approx$ 1.6 mag), as a result of differential reddening and photometric errors. For 7 objects in the Cygnus X area the CMD constrast allowed parameter estimates, for which designations and related nebula are indicated in Table 2 (columns 1 and 2, respectively). The remaining objects are affected by nebulosity, crowding and/or the possibility of too faint member stars.

We measured the colour of the upper Main Sequence (column 3 of Table 2) and the 
magnitude of the 10th brightest star in the cluster (J-K$_s$) strip (column 4 of Table 2). This minimizes field contamination and  stellar evolution effects for massive stars. The objects are assumed to be  very young embedded clusters, but the results
will not change much for relatively evolved clusters up to 6-7 Myr, as long
as  the 10th brightest star belongs to the MS. Indeed, far
infrared and radio observations (Odenwald et al. \cite{ode2}) suggested ages younger 
than 0.1 Myr in DR\,6 and DR\,7, which are predicted to be  powered by late O and early B stars, and  DR\,22 by an O6 ZAMS star. As (J, K$_s$) template we built a 2MASS
CMD of NGC\,6910, which is itself projected on the 
eastern side of Cygnus X. It has an age of 6.5 Myr, a reddening value
E(B-V) = 1.02, an absolute distance modulus (m-M)$_o$ = 11.2 and a distance from 
the Sun {\it d$_{\odot}$} = 1.74 kpc (Delgado \& Alfaro \cite{del}). 

We use the extinction curve by Cardelli et al. (\cite{car}) 
with total-to-selective absorption R$_V$ = 3.1, the K and K$_s$ filter transmissions (Persson et al. \cite{per}) and  A$_K$ = 0.618 E(J-K) (Mathis \cite{mat}) to derive 
the relations  A$_K$/A$_V$ = 0.112, A$_{K_s}$/A$_V$ = 0.118 and A$_{K_s}$=0.670 E(J-K$_s$). The reddening and distance estimates for each cluster were obtained by comparison of the upper Main Sequence colour and 10th brightest star K$_s$ magnitude in the CMDs of the cluster and the template. Column 5 and 6 of Table 2 list the upper Main Sequence colour difference  $\Delta$(J-K$_s$) with respect to the template (in the
sense NGC\,6910 - Object), and the resulting reddening, respectively. The resulting absolute distance modulus and distance are given in columns 7 and 8, respectively. The 10th brightest star in the template cluster NGC\,6910 corresponds to an absolute magnitude M$_{K_s}$ = 0.07. Deeply embedded regions tend to have R$_V$ values higher than the standard one (e.g. Comer\'on \& Torra (\cite{com}) and references therein). Nevertheless the use of infrared bands minimizes reddening uncertainties. The upper Main Sequence colour half width W(J-K$_s$)/2 = 0.8 (Fig. 4) converts to absorption error $\epsilon$(A$_{K_s}$) = $\pm$0.54 which in turn gives a distance error $\approx$ $\pm$0.4 kpc at {\it d$_{\odot}$} $\approx$ 1.5 kpc.

   The Cygnus X clusters in Table 2 are in the range  1.0 $<$ {\it d$_{\odot}$(kpc)} $<$ 
1.8, with an average distance {\it d$_{\odot}$} $\approx$ 1.36 $\pm$0.29 kpc. The values suggest a depth of 800 pc. The resulting linear ratios along major and minor axes (along the sky at the derived average distance), and depth are 3:1:11. The objects
are at a comparable distance or foreground to NGC\,6910, and Cyg\,OB2 also 
estimated to be located at {\it d$_{\odot}$} $\approx$ 1.7 kpc (Massey \& Thompson \cite{mas}). The derived distance range for Cygnus X suggests a more compact distribution than previous estimates, but still larger than the largest OB associations. At least part of it could be consistent  with star formation arising from a local coherent physical event, probably  related to  Cyg\,OB2  and NGC\,6910. Deep photometry for the present clusters (Table 2) and the additional ones in the Cygnus X direction (Table 1),  coupled 
to high quality spectroscopy for the knowledge of spectral types, may shed further light on the depth issue and history of star formation in the area.
   
\begin{table}
\caption[]{Reddening and distance for Cygnus X embedded objects}
\begin{scriptsize}
\label{tab1}
\renewcommand{\tabcolsep}{0.9mm}
\begin{tabular}{lccccccc}
\hline\hline
 Cluster&     Rel. &     (J-K$_s$) &   (K$_s$) & $\Delta$(J-K$_s$)&E(J-K$_s$)& (m-M)$_o$ &  {\it d$_{\odot}$} \\
 &Neb.&uMS&10$^{th}$&&&&(kpc)\\
\hline
NGC6910 &       ---  &       0.45 &    11.65  &        0.00  &        0.56  & 
 11.20 &    1.7 \\
 Obj.6 &        DR9 &        2.50 &    12.95  &        2.05  &        2.61  & 
 11.13 &    1.7\\ 
 Obj.7  &       DR6   &      2.90 &    12.50   &       2.45  &        3.01 & 
 10.41  &   1.2 \\ 
 Obj.8 &        DR13 &       2.60 &    12.80   &       2.15  &        2.71 & 
 10.91  &   1.5 \\
 Obj.11 &       DR7  &       2.55 &    12.10   &       2.10    &      2.66 & 
 10.25 &    1.1\\
 Obj.13 &       DR22 &       3.20 &    12.40  &        2.75  &        3.32 & 
10.11  &   1.0\\
 Obj.15  &      DR17 &       2.40 &    12.10   &       1.95  &        2.51 & 
 10.35 &    1.2 \\
 Obj.21 &       DR16?  &     2.55  &   13.10  &        2.10   &       2.66 & 
 11.25  &   1.8 \\
\hline
\end{tabular}
\end{scriptsize}
\end{table}

\subsection{Galactic Centre}

\begin{figure}
   \centering
 \resizebox{\hsize}{!}
 {\includegraphics{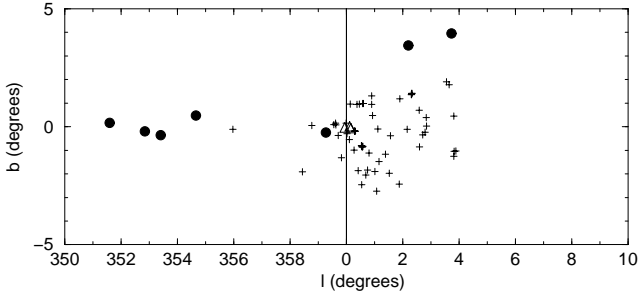}}
\caption{Angular distribution of the present central clusters and candidates (circles), those from Dutra \& Bica (\cite{db20}) (plus signs) and well-known central clusters (triangles).}
\label{FigG}
\end{figure}
 
Table 1 contains 7 new IR clusters or candidates  towards the central areas of the Galaxy.  The angular distribution of the previous literature and present objects is shown in Fig. 5. Most of them are  outside the 5$^{\circ}$ radius region which converts to  $\approx$ 600 pc at the distance of the Galactic centre. Two new objects are located within 600 pc, totaling now 60 cluster
candidates.  The asymmetry of the distribution with respect to the
Galactic Nucleus in Fig. 1 is probably an artifact since the region  356 $<$ $\ell$ ($^{\circ}$) $<$ 359  has
not yet  been  systematically surveyed. Added to the candidates are the well-known central clusters: Arches (Nagata et al. \cite{nag2}, \cite{nag3}), Quintuplet (Nagata et al. \cite{nag}, Glass et al. \cite{gla}) and the Young Nuclear Cluster of He I stars (Krabbe et al. \cite{kra}). Recently, the latter object was suggested to be the remains of a dissolved young cluster (Gerhard \cite{ort}).

\begin{table}
\caption[]{IR Dark nebulae towards the central bulge}
\begin{scriptsize}
\label{tab1}
\renewcommand{\tabcolsep}{0.9mm}
\begin{tabular}{lrrrrrrl}
\hline\hline
IRDN&$\ell$&{\it b}& $\alpha$(2000) & $\delta$(2000)&D & d& Remarks\\
&($^{\circ}$)&($^{\circ}$)&hh:mm:ss &$^{\circ}$:$^{\prime}$:$^{\prime\prime}~$&$^{\prime}$&$^{\prime}$~~&\\
\hline
       1  &        0.25&  -0.47&  17:48:02& -28:58:00&    15&    11&                            related to Sh2-19?\\
       2 &         0.58&  -0.86&  17:50:20& -28:53:00&    19&    14&                            related to Sh2-21?\\
       3 &       358.99&   0.09&17:42:51& -29:45:15 &    3  &   3&\\
       4 &       359.34&   0.29&17:42:54& -29:21:00 &    5 &  2.5&\\
       5 &        359.94&   0.17& 17:44:49 &-28:54:10  &  10  &   8 &related to Sh2-17?\\
\hline
\end{tabular}
\end{scriptsize}
\end{table}
 
 During the present study and Paper I we serendipitously came across with 
prominent dark clouds which are opaque in the $K_s$ band. The clouds are
listed in Table 3. Clouds 1, 2 and 5 
have larger angular dimensions and are possibly  related to intervening H\,II regions. The dust clouds 3 and 4 are compact with small angular dimensions and are thus
candidates to nuclear molecular clouds. Nevertheless, they cannot be ruled out
as nearby foreground globules based solely on the images. At any rate, from the dense dust clouds seen in the $K_s$
band projected onto the bulk of the central bulge/disk stellar population, one
can infer that a new generation of massive clusters may form in the central disk.

 Recently,  Portegies Zwart et al. (\cite{por}) carried out numerical simulations of the dynamical evolution of massive star clusters  within $\approx$200 pc of the Galactic Centre. They concluded that the tidal dissolution time is $\approx$70 Myr, but close to the Galactic centre their projected densities drop below the background density within $\approx$ 20 Myr. They estimated that the region within 200 pc could easily harbour as many as  50 massive clusters. Within a circular region  with r $<$ 1.6$^{\circ}$, or  200 pc assuming a Galactic Center  distance {\it d$_{\odot}$} = 8 kpc (Reid \cite{rei}), there are 31 clusters or candidates  in Fig. 5, thus consistent with the model expectations by  Portegies Zwart et al. (\cite{por}), taking into account observational and theoretical uncertainties.

In the following we discuss in more detail the objects probably related 
    to an edge-on central disk. The region within $|\ell|$ = 1.6$^{\circ}$ 
    and $|b|$ = 0.5$^{\circ}$ (Fig. 5) is now relatively well surveyed 
    considering the 3 well-known central clusters, Paper I and the present 
    study samples. It  converts to an edge-on cylinder of radius r $\approx$ 
    200 pc and distance from the plane $|z|$ = 63 pc, where 17 clusters or 
    candidates are projected. From these objects, candidates 1, 17 
and 31 (Paper I)  are 
    apparently not related to any H\,II region and might be clusters in the 
    density contrast/tidal survival age ranges $\approx$7-20 Myr  or 
    $\approx$ 7-70 Myr. Candidate 58 is possibly related to Sh2-17, 
    and candidates 4, 5 and 6 to Sh2-20. Consequently they are suspected 
 to be intervening objects, and were excluded. The age of the Quintuplet Cluster is $\approx$ 4 Myr (Figer et al. \cite{fig}), and similar ages can be inferred from the properties of  massive stars in the Arches cluster (Lang et al. \cite{lan}) and the Young Nuclear cluster (Gerhard \cite{ort}, Krabbe et al. \cite{kra}). The candidates 26 and 52 are possibly  related to the radio H\,II region Complexes Sgr D and Sgr E (Liszt \cite{lis}), respectively. Candidates 53, 54 and 55 are possibly related to the radio source G359.54+0.18, candidate 56 to 
    G359.7-0.4, and the present object 42 (Table 1) to G359.3-0.3. Assuming an H\,II region duration of $\approx$5 Myr these 10 possible massive clusters imply a formation rate of 2 clusters Myr$^{-1}$ in the central disk of radius 200 
pc.

 If the dark clouds 3 and 4 in Table 3 are able to form a massive cluster in the coming Myr, they would provide the same rate of cluster formation in the central disk as that derived from the ionising clusters above.  
 
\subsection{Objects throughout the disk}  
 
Several new IR clusters are reported in Table 1 for the anticentre region.
In particular 3 new IR clusters are related respectively to the H\,II regions 
Sh2-254, Sh2-256 and  Sh2-258. These H\,II regions form a complex together
with Sh2-255 and Sh2-257, which are lobes of the bipolar H\,II region IC\,2162.
The most prominent IR cluster in the complex Sh2-254/258 was reported  by Hodapp (1994) and it is located between the two lobes. The complex now
contains 4 clusters and a designation IC\,2162 IR cluster or Sh2-255/257 IR cluster is more suitable.  
 
In Table 1 two additional  pairs occur: (i) the IR cluster 33 and the stellar group 34, (ii) the
IR cluster 24 and Sh2-233SE IR cluster (Hodapp \cite{hod}). Two IR clusters are related to optical reflection nebulae: vdBH-RN26 and vdBH-RN43 (van den Bergh \& Herbst \cite{van}).
The former appears to be related to the $\eta$ Carinae Complex, and the second
to the RCW\,38 Complex. They are located in the outskirts of the complexes,
and if physically related to them, they would imply formation of less massive
star clusters, still embedded in their dust, which have not  formed a 
massive ionising star.
 
\section{Concluding remarks}
 
A new sample of 42 IR star clusters, stellar groups and candidates was
found on the 2MASS Atlas, together with 1 bright nebula and 5 dark nebulae.
Most of the  objects are in the Cygnus X area, where a census of embedded 
clusters and stellar groups appears to be significant now, with 25 entries. 
A binary frequency among physical stellar groups of 14\% was obtained in Cygnus X. Distances were estimated for 7 Cygnus X 
clusters,  providing a depth of $\approx$ 800 pc. Such range suggests
line-of-sight projections, but it is not excluded that part
of that is coherently connected to the Cyg\,OB2 formation event and
subsequent interaction with the surrounding medium.
 
 We found 7 new IR clusters and candidates towards the central parts of the
Galaxy. The census of IR clusters and candidates towards these central
parts is $\approx$60 within 600 pc, and  31 within
200 pc. These numbers are within expectations for the number of central massive star clusters in terms of formation rate and tidal survival in the central parts of the Galaxy (Portegies Zwart et al. \cite{por}). Opaque dark clouds in 
the 2MASS K$_S$ images of the central parts of the Galaxy led to molecular
cloud candidates, perhaps able to  form the next generation of massive nuclear star clusters. 

Efforts like the present one to identify and
determine  accurate positions for new embedded star clusters are crucial for observational purposes, since radio HII regions can be so much reddened that in many cases they remain undetectable even in the K band. Positive detections by means  of 2MASS can spare observing time in large telescopes, and can shed more light on the population of young embedded clusters in molecular complexes, and clusters in general which are embedded in the Galactic plane.

\begin{acknowledgements}
We thank the referee Dr. Fernando Comer\'on for interesting remarks.
This publication makes use of data products from the Two Micron All Sky Survey, which is a joint project of the University of Massachusetts and the Infrared Processing and Analysis Center/California Institute of Technology, funded by the National Aeronautics and Space Administration and the National Science Foundation.
We employed  catalogues from CDS/Simbad (Strasbourg) and Digitized Sky Survey images from the Space Telescope Science Institute (U.S. Government grant NAG W-2166) obtained using the extraction tool from CADC (Canada). 
We acknowledge support from the Brazilian institution CNPq.   
\end{acknowledgements}

\end{document}